\documentclass[prd,10pt,aps,showpacs,twocolumn,amsmath,amssymb,nofootinbib,longbibliography]{revtex4-1}

\usepackage{natbib}
\usepackage[colorlinks]{hyperref}
\hypersetup{linkcolor=blue,citecolor=blue}

\usepackage{graphicx}
\usepackage{dcolumn}
\usepackage{bm}
\usepackage{color}
\usepackage{soul}

\usepackage{hyperref}

\def\prd{PRD}
\def\pra{PRA}

\begin{document}

\title{Trirefringence in nonlinear magnetoeletric metamaterials revisited}

\author{V. A. \surname{De Lorenci}}
\email{delorenci@unifei.edu.br}
\affiliation{Instituto de F\'{\i}sica e Qu\'{\i}mica,    Universidade Federal de Itajub\'a, \\
Itajub\'a, Minas Gerais 37500-903, Brazil}

\author{A. L. \surname{Ferreira Junior}}
\email{alexsandref@unifei.edu.br}
\affiliation{Instituto de F\'{\i}sica e Qu\'{\i}mica,    Universidade Federal de Itajub\'a, \\
Itajub\'a, Minas Gerais 37500-903, Brazil}

\author{Jonas P. Pereira}
\email{jonas.pereira@ufabc.edu.br}
\affiliation{Universidade Federal do ABC, Centro de Ci\^encias Naturais e Humanas, Avenida dos Estados 5001- Bang\'u, CEP 09210-580, Santo Andr\'e, SP, Brazil.}
\affiliation{ Mathematical Sciences and STAG Research Centre, University of Southampton, Southampton, SO17 1BJ, United Kingdom.}

\begin{abstract}
Trirefringence is related to the existence of three distinct phase velocity solutions (and polarizations) for light propagation in a same wave-vector direction. This implies that when a trirefringent medium refracts a light ray, it is split into three rays with different velocities and linearly independent polarizations.
Here a previous investigation \cite{2012PhRvA..86a3801D} is revisited and its results are generalized to include a broader class of magnetoelectric materials. Moreover, it is argued that trirefringent media could already be achieved using present day technologies. Examples are given to support that focused on built layered media under the influence of applied electromagnetic fields.
%

\end{abstract}

\maketitle

\section{Introduction}

Metamaterials are extremely important because they hold the promise of huge advancements in optical devices and new technologies \citep{2010opme.book.....C,Zheludev582,2011NaPho...5..523S,2016NatNa..11...23J}. Motivations for them stem from the studies of Veselago with negative dielectric coefficients and the unusual properties they could exhibit \citep{1968SvPhU..10..509V}. Nowadays, metamaterials are an experimental reality and they are broadly understood as man-made media with subwavelength structures such that their dielectric coefficients (and hence their light properties) be controllable \citep{book-metamaterials}. Experimental approaches to metamaterials started with the tailoring of ``meta-atoms'' (for an example of that, see \citet{1999ITMTT..47.2075P}), leading to media with negative refractive indices \citep{2000PhRvL..84.4184S,2001Sci...292...77S,2004PhT....57f..37P}, but now metamaterials have reached an incredible abundance of phenomena and media settings (see for instance \citep{2010opme.book.....C,Zheludev582,2011NaPho...5..523S,2016NatNa..11...23J} and references therein).
%
Metamaterials can also be used to test analogously several gravity models, as for instance \citep{2014Galax...2...72S,2009NatPh...5..687G,2013NaPho...7..902S} and references therein.
%
Through transformation optics \cite{2009PrOpt..53...69L,2010NatMa...9..387C}, recipes for metamaterials with desired light trajectories 
can also be readily obtained.
For the majority of above aspects, focus is put on linear metamaterials, naturally due to the direct relationship of the dielectric coefficients with controllable parameters.
However, nonlinear metamaterials \citep{2014RvMP...86.1093L}
are also rapidly gaining interest, especially for the unique uses they could have \citep{2014RvMP...86.1093L,2012PhRvA..86c3816R,2012ApPhL.101e1103R}.

A special class of nonlinear media on which we will focus in this work is magnetoelectric materials \citep{2005JPhD...38R.123F}.
%
In such media, permittivity (polarization) can depend on the magnetic field and permeability (magnetization) can have an electric field dependence. 
The advantage of metamaterials is that nonlinear magnetoelectric effects there could be created and  controlled. It is already known that they could happen for instance in periodic, subwavelength and polarizable structures, such as split ring resonators \citep{2012ApPhL.101e1103R}. It has also been recently shown that nonlinear magnetoelectric metamaterials could lead to trirefringence \citep{2012PhRvA..86a3801D}, associated with three different solutions to the Fresnel equation \citep{landau1984electrodynamics}. In this work we concentrate on some facets of this phenomenon, such as theory and experimental proposals for it.

Known not to exist in linear media \citep{1969AcOpt..16..133W}, trirefringence would thus be an intrinsically nonlinear effect. Analysis has suggested it might appear in an anisotropic medium presenting some negative components of its permittivity tensor, while its permeability should depend on the electric field (magnetization dependent on the electric field) \cite{2012PhRvA..86a3801D}. Symmetries of Maxwell equations for material media also suggest that trirefringence could take place in media with constant  anisotropic permeability tensors and permittivity tensors depending on magnetic fields (magnetically induced polarization) too. The interest associated with trirefringence is that media exhibiting it could support three linearly independent light polarizations. For each polarization there would be a specific refractive index (phase velocity). Therefore, upon refraction of an incident light ray on a trirefringent medium, it should split into three light rays, each one of them with its specific ray velocity and polarization.
Interestingly, trirefringence might also occur in nonlinear theories of the electromagnetism related to large QED and QCD field regimes \citep{2013PhRvD..88f5015D}. In both cases, though, it is important to stress that trirefringence is an effective phenomenon. The same happens in metamaterials, given that the dielectric coefficients there are clearly effective and just valid for a range of frequencies.

Applications for trirefringence could be thought of regarding the extra degree of freedom brought by the third linearly independent polarization it presents. For instance, if information is stored in polarization modes, then trirefringent media would be $30\%$ more efficient than birefringent media. Another possible application could be related to its intrinsic light propagation asymmetry \citep{2012PhRvA..86a3801D}.
The reason is because trirefringent media lead to three waves propagating in a given range of wave directions (two extraordinary and one ordinary), while the opposite range just has one (ordinary wave) (see Fig. 1 of Ref. \citep{2012PhRvA..86a3801D}).
However, our main motivation in this work is conceptual: we want to argue there might already be feasible ways of having trirefringence with known metamaterials, and hence the possibility of having three linearly independent polarizations to electromagnetic waves in the realm of Maxwell's electrodynamics. In this sense, our analysis is a natural extension of the first ideas presented in \citet{2012PhRvA..86a3801D}.

We structure our work as follows. In Sec. \ref{wave-propagation} we elaborate on light propagation in nonlinear media in the limit of geometric optics. 
Section \ref{trirefringence} is particularly devoted to trirefringence analysis and aspects of the magnetoelectric media which could support it. 
Estimates of the effect as well as proposals of possible physical media where trirefringence could be found is given in Sec. \ref{toy-model-trirefringence}. Particularly, general estimates and graphic visualization of the effect is studied in Sec. \ref{general}, and easy-to-build proposals for trirefringent system based on layered media and estimates of the strength and resolution of the effect are presented in Sec. \ref{estimates}. 
Finally, in Sec. \ref{discussion}, we discuss and summarize the main points raised in this work. 
An alternative method to derive the Fresnel equation from Maxwell's equations and constitutive relations is presented in an appendix.

Unless otherwise specified, we work with international units. 
Throughout the text Latin indices $i, j, k . . .$ run from 1 to 3 (the three spatial directions), and we use the Einstein convention that repeated indices in a monomial indicate summation.

\section{Wave propagation in material media}
\label{wave-propagation}
We start with Maxwell's equations in an optical medium at rest in the absence of free charges and currents \citep{2012PhRvA..86a3801D},
\begin{eqnarray}
\partial_{i}D_{i}&=&0,  \hspace{5mm} \epsilon_{ijk}\partial_{i}E_{j}=-\partial_{t}B_{k},
\label{1}
\\
\partial_{i}B_{i}&=&0,  \hspace{5mm} \epsilon_{ijk}\partial_{i}H_{j}=\partial_{t}D_{k},
\label{2}
\end{eqnarray}
together with the general constitutive relations between the fundamental fields $\vec{E}$ and $\vec{B}$ and the induced ones $\vec{D}$ and $\vec{H}$,
\begin{eqnarray}
D_{i}&=&\varepsilon_{ij}(\vec{E},\vec{B})E_{j}+\tilde{\varepsilon}_{ij}(\vec{E},\vec{B})B_{j},
\label{constitutive1}
\\
 H_{i}&=&\mu_{ij}^{\mbox{\tiny $(-1)$}}(\vec{E},\vec{B})B_{j}+\tilde{\mu}_{ij}^{\mbox{\tiny $(-1)$}}(\vec{E},\vec{B})E_{j}.
\label{constitutive2}
\end{eqnarray}
Here the coefficients $\varepsilon_{ij}$, $\tilde{\varepsilon}_{ij}$, $\mu_{ij}^{\mbox{\tiny $(-1)$}} $, and $\tilde{\mu}_{ij}^{\mbox{\tiny $(-1)$}} $, describe the optical properties of the material.  
%
%

Our interest relies on the study of light rays in a nonlinear magnetoelectric material. Hence, we may restrict our investigation to the domain of the geometrical optics.
We use the method of field disturbance \cite{hadamard1903} and define \cite{delorenci2004}  $\Sigma$, given by $\phi(t,\vec{x})=0$, to be a smooth (differentiable of class $C^{n},n > 2$) hypersurface. The function $\phi(t,\vec{x})$ is understood to be a real valued smooth function and regular in a neighbourhood $U$ of $\Sigma$. The spacetime is divided by $\Sigma$ into two disjoint regions: $U^{-}$, for which $\phi(t,\vec{x})<0$, and $U^{+}$, corresponding to $\phi(t,\vec{x})>0$. 

The step of an arbitrary function $f(t,\vec{x})$ (supposed to be a smooth function in the interior of $U^{\pm}$) through the borderless surface $\Sigma$ is a smooth function in $U$ and is calculated by
\begin{equation}
[f(t,\vec{x})]_{\Sigma}\doteq \underset{{P^{\pm}}\rightarrow P}{\lim} [f(P^{+})-f(P^{-})],
\end{equation}
with $P^{+}$,$P^{-}$ and $P$ belonging to $U^{+}$,$U^{-}$ and $\Sigma$, respectively. The electromagnetic fields are supposed to be smooth functions in the interior of $U^{+}$ and $U^{-}$ and continuous across $\Sigma$ ($\phi$ is now taken as the eikonal of the wave). However, they have a discontinuity in their first derivatives such that  \cite{hadamard1903}
\begin{equation}
[\partial_{t}E_{i}]_{\Sigma}=\omega e_{i}, \hspace{5mm} [\partial_{t}B_{i}]_{\Sigma}=\omega b_{i},
\label{5}
\end{equation}
\begin{equation}
[\partial_{i}E_{j}]_{\Sigma}=-q_{i} e_{j}, \hspace{5mm} [\partial_{i}B_{j}]_{\Sigma}=-q_{i} b_{j},
\label{6}
\end{equation}
where $e_{i}$ and $b_{i}$ are related to the derivatives of the fields on $\Sigma$ and correspond to the components of the polarization of the propagating waves. The quantities $\omega$ and $q_{i}$ are the angular frequency and the $i$-th component of the wave-vector.

Thus, substituting Eqs. (\ref{5}) and (\ref{6}) into the Maxwell equations, together with the constitutive relations, we obtain the eingenvalue equation,
\begin{equation}
Z_{ij}e_{j}=0,
\label{ev}
\end{equation}
where $Z_{ij}$ gives the elements of the Fresnel matrix, and is given by (see the appendix for an alternative way of deriving these results)
\begin{equation}
Z_{ij}=C_{ij}v^{2}+\left(\epsilon_{ikl}\tilde{H}_{lj}+\epsilon_{nkj}\tilde{C}_{in} \right)\hat{q}_{k}v + \\\epsilon_{ikl}\epsilon_{npj}H_{ln}\hat{q}_{k}\hat{q}_{p},
\label{Zij}
\end{equation}
with the definitions
\begin{eqnarray}
C_{ij}&=&\varepsilon_{ij}+\frac{\partial\varepsilon_{in}}{\partial E_{j}}E_{n}+\frac{\partial\tilde{\varepsilon}_{in}}{\partial E_{j}}B_{n},
\label{C}
\\
\tilde{C}_{ij}&=&\tilde{\varepsilon}_{ij}+\frac{\partial\tilde{\varepsilon}_{in}}{\partial B_{j}}B_{n}+\frac{\partial\varepsilon_{in}}{\partial B_{j}}E_{n},
\label{ctilde}
\\
H_{ij}&=&\mu^{\mbox{\tiny $(-1)$}} _{ij}+\frac{\partial\mu^{\mbox{\tiny $(-1)$}} _{in}}{\partial B_{j}}B_{n}+\frac{\partial\tilde{\mu}^{\mbox{\tiny $(-1)$}} _{in}}{\partial B_{j}}E_{n},
\label{h}
\\
\tilde{H}_{ij}&=&\tilde{\mu}^{\mbox{\tiny $(-1)$}} _{ij}+\frac{\partial\tilde{\mu}^{\mbox{\tiny $(-1)$}} _{in}}{\partial E_{j}}E_{n}+\frac{\partial\mu^{\mbox{\tiny $(-1)$}} _{in}}{\partial E_{j}}B_{n},
\label{htilde}
\end{eqnarray}
Furthermore, we have defined the phase velocity as $v=\omega/q$, the unit wave vector as $\hat{q}=\vec{q}/q$, and its $i$-th component as $\hat{q}_{i}$.

\section{Trirefringence in nonlinear magnetoelectric media}
\label{trirefringence}

Magnetoelectric phenomenona in material media are related to the induction of magnetization or polarization (or both) by means applied electric  or magnetic fields, respectively.  In order to obtain the description of these phenomena, we start by expanding the free energy of the material in terms of the applied fields as follows  \citep{2005JPhD...38R.123F},
\begin{eqnarray}
F(\vec{E},\vec{H})=&F_{0}& - P^{S}_{i}E_{i} - M^{S}_{i}H_{i}  
\nonumber
\\
&-& \frac{1}{2}\varepsilon_{0}\chi_{ij}E_{i}E_{j} -\frac{1}{2}\mu_{0}\chi^{\mbox{\tiny $(m)$}} _{ij}H_{i}H_{j} - \alpha_{ij}E_{i}H_{j}
\nonumber
\\
&-& \frac{1}{2}\beta_{ijk}E_{i}H_{j}H_{k}-\frac{1}{2}\gamma_{ijk}H_{i}E_{j}E_{k} + ...\,,
\end{eqnarray}
with $F_{0}$  the free energy of the material in the absence of applied fields, and the other coefficients in this expansion will the addressed in the discussion that follows. 

Differentiation of $F$ with respect to the $E_i$ and $H_i$ fields leads to the polarization and magnetization vectors,
\begin{align}
P_{i}(\vec{E},\vec{H})=&-\frac{\partial F}{\partial E_{i}}   = P^{S}_{i}+\varepsilon_{0}\chi_{ij}E_{j}+\alpha_{ij}H_{j}
\nonumber
\\
&+ \frac{1}{2}\beta_{ijk}H_{j}H_{k}+\gamma_{j(ik)}H_{j}E_{k}+...
\label{Pi}
\\
M_{i}(\vec{E},\vec{H})=&-\frac{\partial F}{\partial H_{i}}   = M^{S}_{i}+\mu_{0}\chi^{\mbox{\tiny $(m)$}} _{ij}H_{j}+\alpha_{ji}E_{j}
\nonumber
\\
&+ \beta_{j(ik)}E_{j}H_{k}+\frac{1}{2}\gamma_{ijk}E_{j}E_{k}+...,
\label{Mi}
\end{align}
where $P^{S}_{i}$ and $M^{S}_{i}$ represent the components of the spontaneous polarization and magnetization, respectively, whose contribution will be suppressed in our discussion. We use the notation that parenthesis enclosing a pair of indices means symmetrization, as for instance $\beta_{j(ik)} = (\beta_{jik}+\beta_{jki})/2$ .

Coefficients $\alpha_{ij}$ and $\gamma_{ijk}$ are responsible for linear and nonlinear electric-field-induced  effects, respectively. However,
we  specialize our discussion to the case of nonlinear magnetic-field-induced effect parametrized by the coefficients $\beta_{ijk}$. Hence,
\begin{eqnarray}
P_{i}&=&\varepsilon_{0}\chi_{ij}E_{j}+\frac{1}{2}\beta_{ijk} H_{j}H_{k},
\label{Pi2}
\\
M_{i}&=&\mu_{0}\chi^{\mbox{\tiny $(m)$}} _{ij}H_{j}+\beta_{j(ik)}E_{j}H_{k}.
\label{Mi2}
\end{eqnarray}
Furthermore, we assume that the linear electric susceptibility $\chi_{ij}$ is described by a real diagonal matrix, and the linear magnetic permeability is isotropic, $\chi^{\mbox{\tiny $(m)$}} _{ij}=\chi^{\mbox{\tiny $(m)$}}  \delta_{ij}$. Losses are being neglected. In the regime of geometrical optics the wave fields are considered to be negligible when compared with the external fields, which we set as $\vec{E}_{ext}=E\hat{x}$ and $\vec{B}_{ext}=B\hat{y}$.
Now, as $B_{i}=M_{i}+\mu_{0}H_{i}$ and $D_{i}=P_{i}+\varepsilon_{0}E_{i}$, we obtain that
\begin{eqnarray}
B_{i}&=&\mu_{0}(1+\chi^{\mbox{\tiny $(m)$}} )H_{i}+\beta_{j(ik)}E_{j}H_{k},
\label{Bi}
\\
D_{i}&=&\varepsilon_{0}(\delta_{ij}+\chi_{ij})E_{j}+\frac{1}{2}\beta_{ijk} H_{j}H_{k}.
\label{Di}
\end{eqnarray}
We restrict our analysis to case where $\beta_{ijk} \ne 0$ only if $j=k$, which implies that $\vec{H}$ is parallel to $\vec{B}$. Additionally, assuming that $\beta_{zyy}$ is negligible and that $\varepsilon_{0}\chi_{xx}E \gg \beta_{xyy}H^{2}$, we obtain that $\vec P = (\varepsilon_{0}\chi_{xx}E,\, \beta_{yyy}H^{2}/2,\, 0)$ and $\vec M = (0,\, \mu_{0}\chi^{\mbox{\tiny $(m)$}} H+\beta_{xyy}EH,\,0)$. Now, returning these results into $\vec D$ and $\vec B$ fields and comparing with the general expressions for the constitutive relations we get that the optical system under consideration is characterized by the following electromagnetic coefficients:
\begin{align}
&\varepsilon_{ij} = \varepsilon_{0}(\delta_{ij}+\chi_{ij}),
\label{eij}
\\
&\tilde\varepsilon_{ij} = \frac{\beta_{yyy}B}{2\mu^{2}} \delta_{ij},
\label{etij}
\\
&\mu_{ij}^{\mbox{\tiny $(-1)$}} =\frac{1}{\mu}\delta_{ij},
\label{mij}
\\
&\tilde{\mu}_{ij}^{\mbox{\tiny $(-1)$}}  =0,
\label{mtij}
\end{align}
where we have defined the nonlinear magnetic permeability
\begin{equation}
\mu \doteq \mu_{0}(1+\chi^{\mbox{\tiny $(m)$}} )+\beta_{xyy}E.
\label{m}
\end{equation}

The Fresnel matrix from Eq.~(8) is now given by
\begin{align}
Z_{ij}=&\left(\varepsilon_{ij}-\frac{\mu'}{\mu^{3}}\delta_{iy}\beta_{yyy}B^{2}E_{j}\right)v^{2}
\nonumber
\\ &+ \left[\frac{\delta_{iy}}{\mu^{2}}\beta_{yyy}(\vec{B}\times\hat{q})_{j}+\frac{\mu'}{\mu^{2}}(\vec{B}\times\hat{q})_{i}E_{j}\right]v 
\nonumber
\\
&- \frac{1}{\mu}\left(\delta_{ij}-\hat{q}_{i}\hat{q}_{j}\right),
\label{zij}
\end{align}
where
\begin{equation}
\mu'\doteq\frac{1}{E}\frac{\partial\mu}{\partial E}=\frac{\beta_{xyy}}{E}.
\label{m'}
\end{equation}
We will be particularly interested in materials presenting a natural optical axis, and set
\begin{equation}
\varepsilon_{ij}=diag(\varepsilon_{\parallel},\varepsilon_{\perp},\varepsilon_{\perp}).
\label{eps}
\end{equation}

Nontrivial solutions of the eigenvalue problem state by Eq.  (\ref{ev}) can be found by means of the generalized Fresnel equation $det\vert Z_{ij}\vert=0$, where $\vert Z_{ij}\vert$ symbolizes the matrix whose elements are given by $Z_{ij}$. This equation can be cast as
\begin{equation}
det\vert Z_{ij}\vert=-\frac{1}{6}(Z_{1})^{3}+\frac{1}{2}Z_{1}Z_{2}-\frac{1}{3}Z_{3}=0,
\label{fresnelleq}
\end{equation}
where we defined the traces $Z_{1}=Z_{ii}$, $Z_{2}=Z_{ij}Z_{ji}$, and $Z_{3}=Z_{ij}Z_{jk}Z_{ki}$.

Now, calculating the above traces and returning into Eq. (\ref{fresnelleq}) we obtain the following quartic equation for the phase velocity of the propagating wave,
\begin{equation}
av^{4}+bv^{3}+cv^{2}+dv+e=0,
\label{quartic}
\end{equation}
where
\begin{align}
a=&6\varepsilon_{\parallel}\varepsilon^{2}_{\perp},
\label{a}
\\
b=&6\frac{\mu'}{\mu^{2}}EB\varepsilon^{2}_{\perp}\hat{q}_{z},
\label{b}
\\
c=&-\frac{6}{\mu}\varepsilon_{\perp}\bigg[\varepsilon_{\parallel}(\hat{q}^{2}_{x}+1) +\varepsilon_{\perp}(\hat{q}^{2}_{y}+\hat{q}^{2}_{z})
\nonumber\\
&- \frac{\mu' \beta_{yyy}}{\mu^{3}}EB^{2}\hat{q}_{x}\hat{q}_{y}\bigg],
\label{c}
\\
d=&-\frac{6}{\mu^3}B\hat{q}_{z}\left[\mu' E\varepsilon_{\perp}+\beta_{yyy}\hat{q}_{x}\hat{q}_{y}(\varepsilon_{\perp}-\varepsilon_{\parallel})\right],
\label{d}
\\
e=&\frac{6}{\mu^{2}}\left[\varepsilon_{\parallel}\hat{q}^{2}_{x}+\varepsilon_{\perp}(\hat{q}^{2}_{y}+\hat{q}^{2}_{z})-\frac{\mu'\beta_{yyy}}{\mu^{3}}EB^{2}\hat{q}_{x}\hat{q}^{2}_{y}\right] .
\label{e}
\end{align}
This result generalizes the study presented in Ref. \cite{2012PhRvA..86a3801D}. Particularly, the term containing the coefficient $\beta_{yyy}$ is linked to the coefficient $\tilde\epsilon_{ij}$ in the constitutive relations, and makes the whole system consistent with the free energy approach used here to derive polarization and magnetization vectors that characterize the medium.

Now, if we set the propagation in the xz-plane, i.e, $\hat{q}= sin \theta \hat{x}+cos \theta \hat{z}$, the phase velocity solutions of Eq. (\ref{quartic}) reduce to the ones studied in Ref. \cite{2012PhRvA..86a3801D}, and are given by
\begin{equation}
v_{o}=\pm\frac{1}{\sqrt{\mu\varepsilon_{\perp}}},
\end{equation}
\begin{equation}
v^{\pm}_{e}=-\sigma\hat{q}_{z}\pm\sqrt{(\sigma\hat{q}_{z})^{2}+\frac{1}{\mu\varepsilon_{\parallel}}\left[\left(\frac{\varepsilon_{\parallel}}{\varepsilon_{\perp}}-1\right)\hat{q}^{2}_x+1\right]}
\end{equation}
where we defined the velocity dimensional quantity $\sigma\doteq {\mu'EB}/{2\mu^{2}\varepsilon_{\parallel}}$.
Here, the solution $v_{o}$ does not depend on the direction of the wave propagation and is referred to as the ordinary wave, whereas  $v^{\pm}_{e}$ depend on the direction of the wave propagation, and are called extraordinary waves.

Now, the polarization modes $\vec{e}$ corresponding the above described wave solutions are given by the eigenvectors of Eq. (\ref{ev}). Introducing $Z_{ij}$ from Eq. (\ref{zij}) into the eigenvalue equation, we get the following equation relating the components of $\vec{e}$,
\begin{widetext}
\begin{align}
&\left[\frac{\varepsilon_{\parallel}}{\varepsilon_{\perp}}v^{2} +   2\mu\varepsilon_{\parallel}\sigma v_{o}^{2}\hat{q}^{2}_{z} v   -\left(\frac{\varepsilon_{\parallel}}{\varepsilon_{\perp}}\hat{q}^{2}_{x}+q^{2}_{z}\right)v^{2}_{o}\right]e_{x}=0,
\label{first}
\\
&\left[-2\mu\varepsilon_{\parallel}\sigma v^{2}+\left(\hat{q}_{z}+\frac{\varepsilon_{\parallel}\hat{q}^{2}_{x}}{\varepsilon_{\perp}\hat{q}_{z}}\right)v\right]\frac{\beta_{yyy}B}{\mu}e_{x}+\left(\frac{v^{2}}{v^{2}_{o}}-1\right)e_{y}=0,
\label{second}
\\
&\left(\mu\varepsilon_{\perp}v^{2}-\hat{q}^{2}_{x}\right)e_{z} - (2\mu\varepsilon_{\parallel}\sigma v-\hat{q}_{z})\hat{q}_{x}e_{x}=0.
\label{third}
\end{align}

Note that the coefficient of $e_x$ in Eq. (\ref{first}) is zero only when $v=v^{\pm}_{e}$. Therefore, straightforward calculations lead us to the polarization vectors  $\vec{e}_{o}$ and $\vec{e}^{\;\pm}_{e}$ corresponding to the ordinary and extraordinary waves, respectively,
\begin{align}
\vec{e}_{o}=&\hat{y},
\label{pe0}
\\
\vec{e}^{\;\pm}_{e}=&N^{\pm}\bigg\lbrace\hat{x}-\frac{\varepsilon_{\parallel}\beta_{yyy}Bv^{\pm}_{e}}{\mu\varepsilon_{\perp}}\bigg[1 
+\frac{1}{(\varepsilon_{\perp}/\varepsilon_{\parallel}-1)\hat{q}^{2}_{z}-2\mu\varepsilon_{\perp}\sigma\hat{q}_{z}v^{\pm}_{e}}\bigg]\hat{y}-\frac{\varepsilon_{\parallel}\hat{q}_{x}}{\varepsilon_{\perp}\hat{q}_{z}}\hat{z}\bigg\rbrace,
\label{peext}
\end{align}
where $N^{\pm}$ holds for the normalization constants related to the extraordinary modes.
\end{widetext}

As we see, there will be up to three distinct polarization vectors, depending on the phase velocity solutions. In a same direction there will be only one possible solution for the ordinary wave, as they will always have opposite signs. On the other hand, it is possible to exist two distinct solutions for the extraordinary waves in a same direction, i.e., presenting the same signal. The condition behind this possibility can be expressed as follows \cite{2012PhRvA..86a3801D},
\begin{equation}
-1<\frac{1}{\mu\varepsilon_{\parallel}\sigma^{2}}\left(\frac{\varepsilon_{\parallel}}{\varepsilon_{\perp}}\tan^{2}\theta+1\right)<0.
\label{tri}
\end{equation}
However, in order to guarantee the existence of the ordinary wave solution we should set $\mu\varepsilon_{\perp} > 0$, which implies that the above condition requires that  $\varepsilon_{\parallel}=-\epsilon_{\parallel} < 0$. Hence, the range of $\theta$ for which trirefringence occurs is such that
$1 >(\epsilon_{\parallel}/\varepsilon_{\perp}) \tan^{2}\theta > 1-\epsilon_{\parallel}\mu\sigma^{2}$.

\section{Estimates for trirefringence}
\label{toy-model-trirefringence}

\subsection{General estimates and pictorial analysis of the effect}
\label{general}
As we have seen, in the particular scenario here investigated, trirefringence occurrence is constrained by the condition set by Eq. (\ref{tri}), which requires a material presenting one negative component of the permittivity tensor. This behavior can naturally be found in plasmonic materials. Furthermore, artificial materials constructed with wire structures are known to follow this behavior for an adjustable range of frequencies. Nevertheless, an estimate of the effect can still be presented in terms of the magnitude of certain coefficients found in natural materials. It is expected that such results can be similarly produced in a tailored material.

Measurements  of magnetoelectric polarization in some crystal systems \cite{liang2011,begunov2013,kharkovskiy2016} have shown that values of the $\beta$ coefficient as larger as $\beta_{xyy}\approx 10^{-16}{\rm sA^{-1}}$ can be found, as for instance in ${\rm HoAl}_3({\rm BO}_3)_4$, for which magnetic fields up to about 10T was applied.  
%
Thereby, the angular range for which trirefringence may occur can be presented as
\begin{eqnarray}
&1-\frac{1.42\times10^{-2}}{(\epsilon_\parallel/\varepsilon_0)(\mu/\mu_0)^3}\left(\frac{\beta_{xyy}}{10^{-16} \rm{sA^{-1}}}\right)^2\left(\frac{B}{10 \rm{T}}\right)^2
< \frac{\epsilon_\parallel}{\varepsilon_\perp}\tan^2{\theta} < 1. \notag \\
  & \label{range}
\end{eqnarray}
In the above result we can approximate $\mu/\mu_0 \approx 1+ \chi^{\mbox{\tiny $(m)$}}$, as higher order term in $\beta$ can be neglected.
The permittivity coefficients can be expressed in terms of linear susceptibilities as $\varepsilon_{\perp}=\varepsilon_0(1+\chi_{22})$ and $\epsilon_{\parallel}=\varepsilon_0(\vert\chi_{11}\vert-1)$.

If we assume typical values for the permittivity coefficients as $\epsilon_{\parallel}/\varepsilon_0 = 1.5$ and  $\varepsilon_{\perp}/\varepsilon_0 = 2$, we obtain $1>0.75\tan^2{\theta} >1-9.4857\times10^{-3}(\mu/\mu_0)^{-3}\left(\beta_{xyy}/10^{-16} \rm{sA^{-1}}\right)^2\left(B/10 \rm{T}\right)^2$. Depending on the magnetic properties of the material the above range can be large enough to be experimentally measured. 
Just to have a numerical result, assuming the above results and approximating $\mu/\mu_0 \approx 1$, we get $0.85471 < \theta(\mbox{rad}) < 0.85707$, i.e., there will be a window of about $0.0024\, \mbox{rad}$  ($\approx 0.14$ degrees) where the effect can be found.
For instance, if we set the direction of propagation at $\theta = 0.8570\, \mbox{rad}$, the ordinary and  extraordinary velocities result $v_{0} \approx 0.707c$, $v_e^+ \approx 8.04\times10^{-4} c$ and $v_e^- \approx 0.103 c$, each of which presenting a distinct polarization vector, as given by Eqs. (\ref{pe0}) and (\ref{peext}).

\begin{figure}[h]
\includegraphics[scale=0.54]{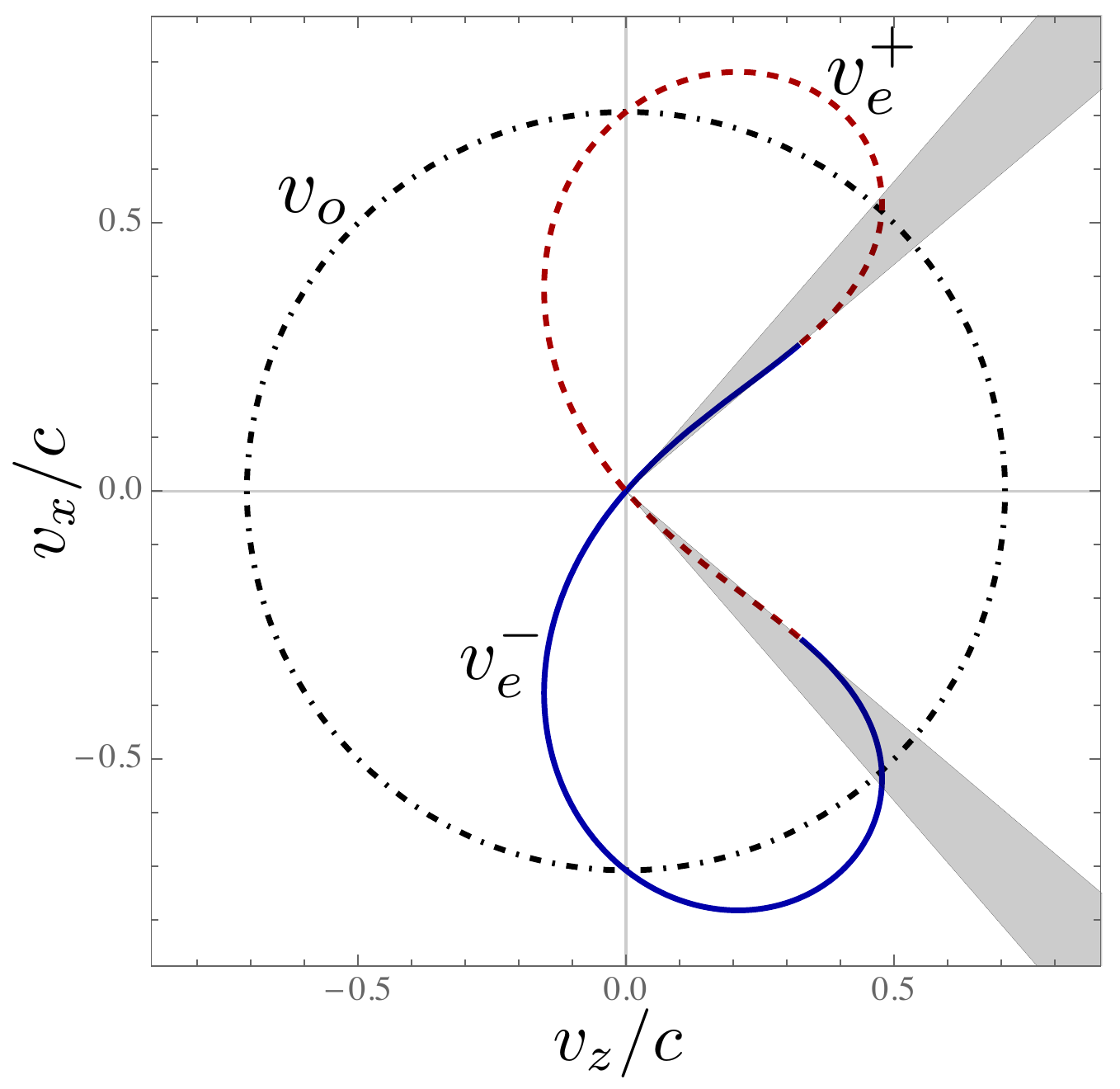}
\caption{Normal surfaces \cite{landau,born} of the nonlinear magnetoelectric medium characterized by Eqs. (\ref{m}) and (\ref{eps}). 
%
%
Here trirefringence phenomenon occurs in the shaded angular sectors.  The solution for the phase velocity of the ordinary wave is depicted by the circular line (labeled by $v_o$), and the extraordinary waves are represented by solid and dashed curves (labeled by $v_e^+$ and $v_e^-$).}
\label{fig1}
\end{figure}

In order to have a better visualization of the effect, let us assume that a system for which $\beta = 10^{-15}sA^{-1}$ can be found (or tailored), and assume an applied magnetic field of 7T. In this case, keeping all the other assumptions, the angular opening for which the effect occurs increases to about $0.1556\, \mbox{rad}$ ($\approx 8.92$ degrees), within $0.70144 < \pm\theta(\mbox{rad}) <  0.85707$, as depicted in Fig. \ref{fig1}.
Note that the trirefringent region is symmetric with respect to the $Z$ direction. This is a consequence of the fact that $v_e^\pm(-\theta)  = - v_e^\mp(\theta)$, i.e.,  the role of the extraordinaries solutions exchanges when $\theta \rightarrow -\theta$. 

\begin{figure}[h]
\includegraphics[scale=0.51]{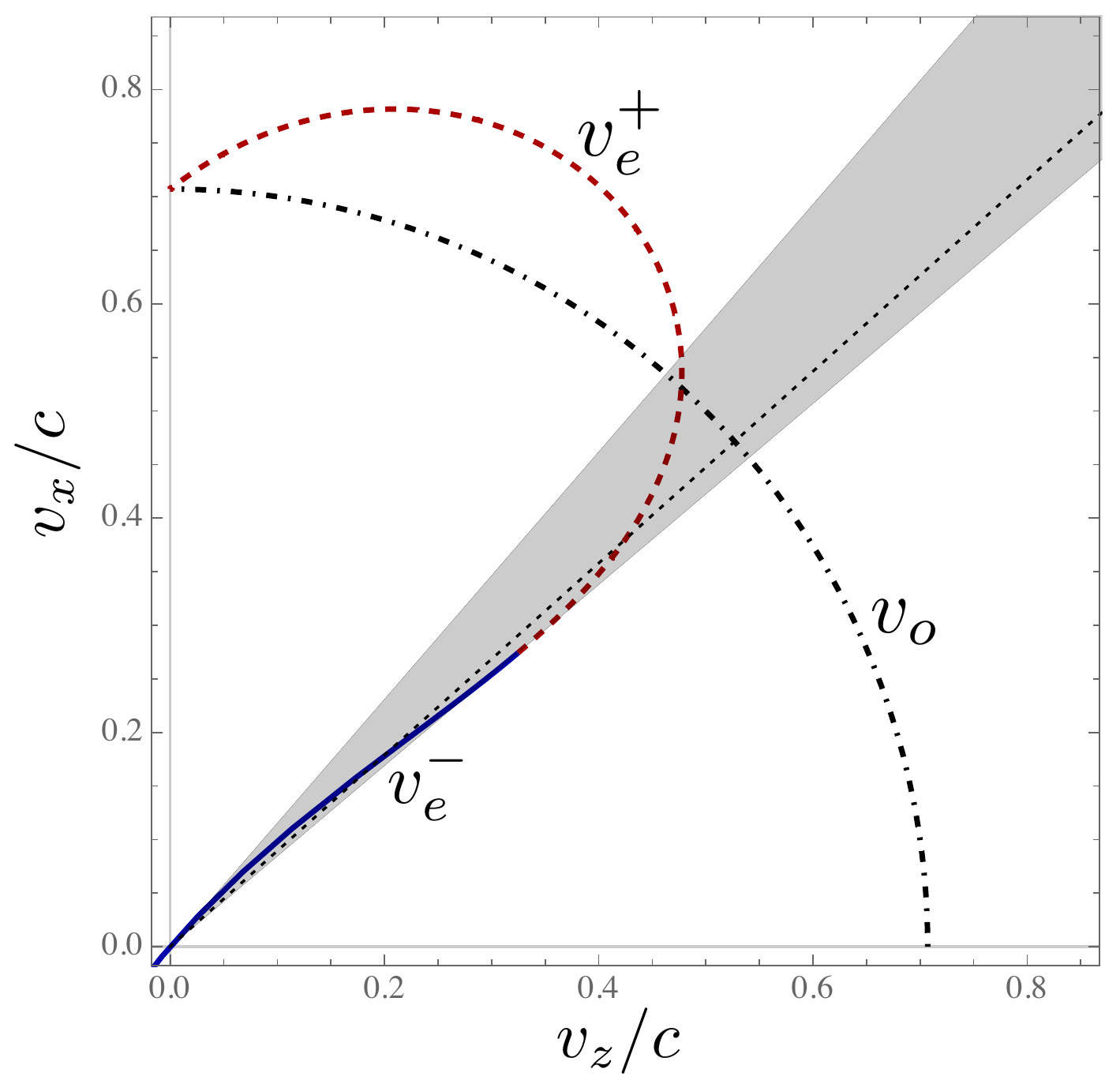}
\caption{Here the dotted straight line represents a specific direction of propagation inside the trirefringence region. It can be clearly seen that in this direction there exist three distinct solutions for wave propagation, which are given by the intersection points between the straight line and the normal surfaces.}
\label{fig2}
\end{figure}

If we concentrate only on the first quadrant of Fig.~\ref{fig1}, and choose a direction of propagation inside the trirefringent window, we clearly see that in such direction there will be three possible solutions -- one ordinary wave solution and two extraordinary ones. This aspect is explored in Fig.~\ref{fig2}, where we have selected the direction $\theta = 0.730\, \mbox{rad}$ ($\approx 41.8^{\circ}$).  
Notice that the dotted straight line meets the normal surfaces in three distinct points, corresponding to the three distinct solutions of wave propagation in that direction. 

Another aspect unveiled by Fig.~\ref{fig2} is the existence of wave solutions for one of the extraordinary rays ($v_e^-$ in the plot)  presenting a phase velocity that can be arbitrarily close to zero. As we see, $v_e^-$ goes to zero as we approach the maximum angle for which trirefringence occurs. This issue is further addressed in the final remarks.

\subsection{Estimates based on a possible multilayered system}
\label{estimates}
Now, we elaborate on a specific proposal for a trirefringent medium. From the above estimate one learns that, ordinarily, trirefringence would be limited to small angular regions because the expected $\beta_{xyy}$s are small. However, Eq. \eqref{range} already suggests a way to increase the angular aperture trirefringence would take place. One should simply choose a small enough $\epsilon_{\parallel}$. Fortunately, this can be achieved with the effective dielectric coefficients in layered media, which are relatively easy to tailor and hence potential experimental candidates to our proposal.

Assume a layered medium whose constituents are two materials with homogeneous dielectric coefficients $(\epsilon_1,\mu_1)$ and $(\epsilon_2,\mu_2)$. For definiteness, take medium ``1'' as an ordinary dielectric material and medium ``2'' as a metal-like material. If one (indefinitely) juxtaposes alternating layers of medium 1 with thickness $d_1$ and medium 2 with thickness $d_2$ such that the directions perpendicular to the thicknesses of the layers are ideally infinite, then one has an idealized layered medium \cite{2006PhRvB..74k5116W}. A clear illustration of what has been just described can be seen in Ref. \cite{2006PhRvB..74k5116W}. If one defines a coordinate system such that the $x$-axis is in the direction of the alternating layers, then the principal effective permittivity components of this layered medium are \cite{2006JOSAB..23..498W}
\begin{equation}
\epsilon_y=\epsilon_z=\frac{\epsilon_1+\eta \epsilon_2}{1+\eta}\label{epsilon-parallel}
\end{equation}
and
\begin{equation}
    \frac{1}{\epsilon_x}=\frac{1}{1+\eta}\left(\frac{1}{\epsilon_1} +\frac{\eta}{\epsilon_2}\right)\label{epsilon-perpendicular},
\end{equation}
where $\eta\equiv d_2/d_1$. Identical expressions also hold for the anisotropic components of the effective permeability tensor of the layered medium \citep{2006PhRvB..74k5116W}. One clearly sees from Eqs. \eqref{epsilon-parallel} and \eqref{epsilon-perpendicular} that when $\eta \rightarrow 0$, the effective dielectric parameters approach $\epsilon_1$, while they are $\epsilon_2$ when $\eta\rightarrow \infty$. This is exactly what one expects because, for instance, when $d_2\rightarrow 0$ ($\eta\rightarrow 0$) the layered medium would basically be medium 1. As of importance in the sequel, if one chooses a particular value for $\epsilon_x$, from Eq. \eqref{epsilon-perpendicular} with free $\epsilon_1$, it implies that
\begin{equation}
    \epsilon_2=\frac{\epsilon_x \eta \epsilon_1}{\epsilon_1(1+\eta) -\epsilon_x}\label{epsilon_metal}.
\end{equation}

Consider now the layered medium is under the influence of applied (controllable) fields such that the electric field is in the $x$-direction and the magnetic field is in the $y$-direction. From our previous notation with regard to trirefringence, we have that $\epsilon_x\equiv \varepsilon_{\parallel}\equiv-\epsilon_{\parallel}<0$ and $\epsilon_y\equiv \varepsilon_{\perp}$. From Eq. \eqref{epsilon_metal}, it thus gets clear that a layered medium under applied fields in our model could just be trirefringent if $\epsilon_2$ is negative. Given that we have chosen medium 2 as a metal-like medium, $\epsilon_2<0$ could always be the case if the frequency $\omega$ of the propagating waves are smaller than the plasma frequency $\omega_p$ of the material. Indeed, the real (R) part of $\epsilon_2$ for metal-like structures is given by (modified Drude-Lorentz model \citep{2010opme.book.....C})
\begin{eqnarray}
\epsilon_2^R=\epsilon_m(\infty)-\epsilon_0\left(\frac{\omega_p}{\omega}\right)^2\label{permittivity-metal},
\end{eqnarray}
where $\epsilon_m(\infty)$ is a positive constant (changing from material to material) \citep{2010opme.book.....C}. Therefore, for any value of $\epsilon_x=-\epsilon_{\parallel}$, one can always find an associated $\omega$ fulfilling it from Eqs. \eqref{epsilon_metal} and \eqref{permittivity-metal}.

In general, losses attenuate electromagnetic waves and they could be estimated by the imaginary (I) parts of the dielectric coefficients. If one takes $\epsilon_{1}$ far from resonance, which we assume here, then its imaginary part is negligible \cite{2010opme.book.....C}. In this case, losses would just be associated with $\epsilon_2$ and for metal-like structures they are of the form $\epsilon^I_2=\epsilon_0\Gamma \omega_p^2/\omega^3$, with $\Gamma$ the damping constant and usually $\Gamma \ll \omega_p$ \cite{2010opme.book.....C}. Simple calculations from Eq. \eqref{epsilon-perpendicular} tell us that
\begin{equation}
    \varepsilon_{\parallel}^I= \frac{\eta (\eta + 1)\epsilon_2^I(\epsilon_{1}^R)^2}{(\epsilon_2^I)^2 + (\epsilon_2^R+\eta \epsilon_{1
}^R)^2}\label{losses}.
\end{equation}
Since small values of $\varepsilon_{\parallel}$ would be interesting for trirefringence, from Eq. \eqref{epsilon_metal} that implies $\epsilon_2^R$ should be small as well and by consequence frequencies of interest should be close to the plasma frequency [see Eq. \eqref{permittivity-metal}]. Then, metal losses should also be small. If one assumes that $\eta \gg 1$, then  $\varepsilon_{\parallel}^I\approx \epsilon_2^I$. Losses in layered media would be negligible when the real parts of their dielectric coefficients are much larger than their imaginary parts. For $|\varepsilon_{\parallel}^R|\gg |\varepsilon_{\parallel}^I|$, from the above, one would need $|\epsilon_{2}^R|\gg |\epsilon_{2}^I|$. $\varepsilon_{\perp}^R\gg \varepsilon_{\perp}^I$ is always the case when frequencies are near the plasma frequency and far from resonant frequencies of dielectric media, exactly what we have assumed in our analysis.

Another important condition for a trirefringent medium would be a nonlinear magnetization induced by an electric field. 
For aspects of its permittivity, we have taken $\eta \gg 1$. Thus, if the permeability of the constituent media are close to a given constant, convenient choices for $\eta$ will render effective coefficients such that $\mu_x\approx \mu_y=\mu_z \approx \mu_2$. In order to have the desired nonlinear response, one should just take a metal-like constituent medium whose permeability is naturally nonlinear. This could be achieved if medium 2 was a ``metametal'', for example a metal array (wire medium \cite{2010opme.book.....C}) of split ring resonators embedded in a given host medium. In addition to the nonlinear response from the split ring resonators \cite{2010opme.book.....C}, such media could also have controllable plasma frequencies \cite{2010opme.book.....C}, which could be set to be in convenient regions of the electromagnetic spectrum, such as the microwave region \cite{2010opme.book.....C}. (Good conducting metals have plasma frequencies usually in the near-ultraviolet and optical \cite{2010opme.book.....C}.) In this case, larger structures could be tailored still behaving as an effective continuous medium, which is of experimental relevance. Moreover, adaptations of split ring resonators could also lead to effective nonlinear responses. Structures such as varactor-loaded or coupled split ring resonators already demonstrate such response in the microwave frequency \cite{2012ApPhL.101e1103R} and hence could also be used in magnetoelectric layered media.

Let us make some estimates for relevant parameters of our trirefringent layered media. Take $\epsilon_{\parallel}/\epsilon_0=1.5\times 10^{-2}$, and keep the other parameters as in the above estimate. For this case, $0.053<7.5\times 10^{-3} \tan^2\theta<1$, or $69.4<\theta(\mbox{degree})<85.1$ [$1.21 <\theta  (\mbox{rad})<1.44$], which is considerably larger than the angular opening of the previous estimate. For typical metal-like parameters, $\Gamma/\omega_p\approx 10^{-3}-10^{-2}$ \cite{2010opme.book.....C}, which means losses could indeed be small for extraordinary waves and rays. If $\beta_{xyy}$ turns out to be smaller than the values estimated, then larger external magnetic fields could be used to increase the angle apertures. Smaller values of $\epsilon_{\parallel}$ could also work but losses in these cases should be investigated with more care, due to the relevance of dispersive effects.


\section{Final remarks}
\label{discussion}

We have shown that combinations of already known nonlinear magnetoelectric materials are feasible media for the occurrence of trirefringence. This sort of effect was previously reported \cite{2012PhRvA..86a3801D} in a more restrictive model, and it has been generalized here. Particularly, when we restrict the propagation of light rays to the xz-plane, the same solutions for phase and group velocities are found. As clearly seen from Eqs. \eqref{pe0} and \eqref{peext} for light rays propagating in the xz-plane, we note that the polarization vectors do not lie on this plane because they explicitly depend on an extra $\beta$ coefficient with regard to the velocities. As an important consequence thereof, the three polarizations of a trirefringent medium are linearly independent. We should also note that in the same way as it occurs in the idealized case \cite{2012PhRvA..86a3801D}, there are three distinct group velocities for each direction of the wave vector in the region where trirefringence takes place. This is an important result because it demonstrates that, analogously to the birefringent case, as a light ray enters a trirefringent slab it splits into three rays with linearly independent polarizations, and each one will follow the trajectory imposed by Snell's law.

We have also given estimates and examples of possible trirefringent media. They have been based on layered media, whose tailoring is relatively easy experimentally. In order to have all ingredients for trirefringence, convenient fields should be applied on these media and one of their constituent parts might have an intrinsic nonlinear magnetic response. This could be achieved, for instance, with arrays of split ring resonators in a given host medium, or any other metallic-meta-atoms which present nonlinear magnetic responses. The trirefringent media proposed would also have controllable plasma frequencies, which would be useful for practical realizations since all the structures should have subwavelength sizes. With layered media one could easily control allowed values for the permittivity components, even render some of them close to zero with small resultant losses. This is an important aspect for trirefringence, given that the set of angles where it takes place would generally depend on the inverse of the permittivity.

Closing, we should mention that the model investigated in this work also allows for the possibility of exploring slow light phenomena. This aspect can be understood by examining Eq.~(\ref{tri}), which sets the condition for the occurrence of trirefringence, or directly by inspecting the specific example explored in Fig.~\ref{fig2}.  The angular opening for which this effect occurs is characterized by two critical angles, the minimum one, $\theta_{\tt min}$, that depends on the magnetoelectric properties of the medium, and the maximum one, $\theta_{\tt max}$, that depends only on the permittivity coefficients. We particularly see that  $v_{e}^{-}(\theta_{\tt max}) = 0$, which means that the velocity of one of the extraordinary rays continually decreases to zero as we move from the trirefringent region to a birefringent region. The whole picture is clearly exposed in Fig.~\ref{fig2}. If we start with $\theta < \theta_{\tt min}$ there will be only one solution for wave propagation, which corresponds to the ordinary wave. When we achieve $ \theta_{\tt min}$ an extraordinary ray will appear, and this particular direction will exhibit birefringence. In fact, at this direction both extraordinary wave solutions coincide. Directions such that $\theta_{\tt min} < \theta < \theta_{\tt max}$ present three distinct solutions as discussed before, characterizing a trirefringent domain. However, as $\theta \rightarrow \theta_{\tt max}$ we have that $v_{e}^{-}(\theta) \rightarrow  v_{e}^{-}(\theta_{\tt max}) = 0$. Hence, we can produce an extraordinary solution with a phase velocity as closer to zero as we wish just by adjusting the direction of propagation as closer to $\theta_{\tt max}$ as possible.  

\begin{acknowledgments}
This work was partially supported by the Brazilian research agencies CNPq (Conselho Nacional de Desenvolvimento Cient\'{\i}fico e Tecnol\'ogico) under Grant No. 302248/2015-3, and FAPESP (Funda\c{c}\~ao de Amparo \`a Pesquisa do Estado de S\~ao Paulo) under grants Nos. 2015/04174-9 and 2017/21384-2.
\end{acknowledgments}

\section{Appendix}
Alternatively to the method of field disturbances, proposed by Hadamard \cite{hadamard1903,papapetrou1974,boillat}, one can derive the Fresnel equation [Eq.~(\ref{Zij})] directly through Maxwell equations [Eqs.~(\ref{1}) and (\ref{2})],  and the constitutive relations [Eqs.~(\ref{constitutive1}) and (\ref{constitutive2})], together with some assumptions on the fields.
The field can be decomposed into probe field $\vec{E}^{{}^{p}}$ and background field $\vec{E}^{{}^{o}}$, $\vec{E}=\vec{E}^{{}^{p}}+\vec{E}^{{}^{o}}$. Where $\vert \vec{E}^{{}^{o}} \vert \gg \vert \vec{E}^{{}^{p}} \vert$, $\vert \nabla\cdot\vec{E}^{{}^{o}} \vert \ll \vert \nabla\cdot\vec{E}^{{}^{p}} \vert$ and $\vert \partial_{t}\vec{E}^{{}^{o}} \vert \ll \vert \partial_{t}\vec{E}^{{}^{p}} \vert$. Note that
\begin{equation}
\frac{\partial}{\partial E_{j}}=\frac{\partial}{\partial E^{{}^{o}}_{n}}\frac{\partial E^{{}^{o}}_{n}}{\partial E_{j}}+\frac{\partial}{\partial E^{{}^{p}}_{n}}\frac{\partial E^{{}^{p}}_{n}}{\partial E_{j}}\simeq \frac{\partial}{\partial E^{{}^{o}}_{j}}.
\nonumber
\end{equation}

The same holds for the magnetic field: $\vec{B}=\vec{B}^{{}^{p}}+\vec{B}^{{}^{o}}$, and so on.

We assume plane wave solutions for the electric and magnetic probe field modes as $\vec{E}^{{}^{p}}=\vec{e}\;\mbox{exp}[i(\omega t-\vec{q}\cdot\vec{r})]$ and $\vec{B}^{{}^{p}}=\vec{b}\;\mbox{exp}[i(\omega t-\vec{q}\cdot\vec{r})]$, from which,
\begin{align}
&\partial_{t} E^{{}^{p}}_{j}=i \omega E^{{}^{p}}_{j}, \hspace{2mm} \partial_{t} B^{{}^{p}}_{j}=i \omega B^{{}^{p}}_{j},
\nonumber
\\
&
\partial_{i} E^{{}^{p}}_{j}=-i q_{i} E^{{}^{p}}_{j}, \hspace{2mm} \partial_{i} B^{{}^{p}}_{j}=-i q_{i} B^{{}^{p}}_{j}.
\nonumber
\end{align}
Substituting these relations in Faraday law [Eq. (\ref{1})] yields $\omega B^{{}^{p}}_{k}=\epsilon_{ijk}q_{i}E^{{}^{p}}_{j}$. Finally, returning this result in Eq. (\ref{2}), we obtain after some algebra the Fresnel equation $Z_{ij}E^{{}^{p}}_{j}=0$, where $Z_{ij}$ is the same as obtained in Eq. (\ref{Zij}).

\bibliography{trirefringence}

\end{document}